\documentclass{PoS}

\newcommand{\bea}{\begin{eqnarray}}
\newcommand{\eea}{\end{eqnarray}}
\newcommand{\be}{\begin{equation}}
\newcommand{\ee}{\end{equation}}

\def\G{\Gamma}

\title{Slavnov-Taylor Identity for the Effective Field Theory of the Color Glass Condensate}

\ShortTitle{ST Identity for the Color Glass Condensate Effective Field Theory}

\author{D.~Binosi\\
        European Centre for Theoretical Studies in Nuclear Physics and Related Areas (ECT*)\\ and
Fondazione Bruno Kessler, Trento, Italy\\
        E-mail: \email{binosi@ectstar.eu}}

\author{\speaker{A.~Quadri}
        \\
        Univ. degli Studi di Milano \& INFN, Sezione di Milano\\
        via Celoria 16, I-20133 Milano, Italy\\
        E-mail: \email{andrea.quadri@mi.infn.it}}

\author{D.~N.~Triantafyllopoulos \\
        European Centre for Theoretical Studies in Nuclear Physics and Related Areas (ECT*)\\ and
Fondazione Bruno Kessler, Trento, Italy\\
        E-mail: \email{trianta@ectstar.eu}}

\abstract{We show that a powerful Slavnov-Taylor (ST) identity exists
for the Effective Field Theory (EFT) of the Color Glass Condensate (CGC), 
allowing to control by purely algebraic means the full dependence 
on the background fields of the fast gluon modes, 
as well as the correlators of the quantum fluctuations 
of the classical gluon source.
We use this formalism to study the change of the background fast
modes (in the Coulomb gauge), induced by the quantum corrections of
the semi-fast gluons. We establish the evolution equation for the
EFT of the CGC, which points towards an algebraic derivation of the
JIMWLK equation. 
Being based on symmetry-arguments only, the approach can be used 
to extend the analysis to arbitrary gauges and to higher
orders in the perturbation expansion of the EFT.}

\FullConference{The European Physical Society Conference on High Energy Physics -EPS-HEP2013\\
		18-24 July 2013\\
		Stockholm, Sweden}

\begin{document}

\section{Introduction}

At high energies and high gluon density the regime of gluon
saturation in the small $x$ region (where $x$ denotes
the longitudinal momentum fraction of the parton) is well
described by the effective field theory (EFT) of the Color
Glass Condensate (CGC)~\cite{Gelis:2010nm}.

A well-known application of the CGC formalism is
the study of heavy ion collisions~\cite{Gelis:2010nm}-\cite{Lappi:2010ek}, where gluon
saturation plays a prominent role in the determination of 
the initial wavefunction of the projectile and target nuclei
as well as in the early phases of the collision.

The EFT of the CGC allows to derive an evolution equation
for the wavefunction of a hadron (or a nucleus) to arbitrarily small
$x$ values. This evolution is controlled by a renormalization
group (RG) equation, the JIMWLK equation~\cite{jimwlk}, that makes it possible to resum
the large logarithmic corrections induced by the radiation
of quantum gluons in a dense environment.

In the CGC picture, the fast colour sources $\rho$
are described by classical configurations associated with
a probability distribution $W_\Lambda[\rho]$ at a given
energy scale $\Lambda$.
The colour sources generate through the Yang-Mills equations of motion
a set of classical (background) gauge fields, describing the
fast gluon modes in the infinite momentum frame, i.e. those wtih momentum component $p^+$ above $\Lambda$.

Quantum gluons are then radiated off the background fast configurations.
In the EFT approach, one needs to integrate the semifast gluon modes with
momentum between $b\Lambda$ and $\Lambda$, where $b$ is a small parameter such that $\alpha_s \ln \frac{1}{b} \ll 1$ (so that a weak coupling expansion
is reliable, despite the fact that the classical field configurations are strong, being of the order $\sim \frac{1}{g}$, and will be resummed).

Then one finds that, in the leading logarithmic approximation, the radiative
corrections amount to the redefinition of the probability distribution
$W_{b\Lambda}[\rho]$, which now describes the CGC at the scale $b \Lambda$.
The classical Yang-Mills equations of motion, relating the
classical fast gluons to the colour sources $\rho$ at the new scale,
are not modified by the quantum evolution. 

The flow from the original probability distribution $W_\Lambda$ to the new
one $W_{b\Lambda}$ is generated by an effective Hamiltonian~\cite{jimwlk,Hatta:2005rn}. 
The resulting functional Fokker-Planck equation for 
the weight function $W_\Lambda[\rho]$ provides the appropriate
mathematical description for the 
one-step quantum evolution in the CGC framework.

It turns out that these results can be traced back to the symmetry content of QCD (in the presence of external gluon field configurations), namely to the Slavnov-Taylor (ST) identity of the theory.
In particular, the deformation of the background fast gluon fields, induced by the integration of the semifast modes, is controlled by the ST identity.
This result extends the one given in~\cite{Binosi:2011ar}-\cite{Binosi:2012st}, where it is shown that
the quantum deformation of the background field can be obtained
through a canonical transformation whose form is dictated by the ST identity.
Moreover, the ST identity also controls the quantum deformation of the Yang-Mills equations of motion for the fast background gluons. 

By exploiting these constraints it is possible to obtain an evolution equation of the effective action (after integration of the semifast modes) with respect to (w.r.t.) the parameter $b$. This holds in a rather general setting and points towards a purely algebraic derivation of the JIMWLK evolution equation.

\section{The classical theory of the CGC}

The Yang-Mills (YM) action in the presence of a colour source $\rho$ is
\begin{eqnarray}
S_{\rm CGC}[A,\rho] = S_{\rm YM}[A]+ S_{W}[A,\rho] \, .
\end{eqnarray}
The conventional Yang-Mills action is given by
\bea
S_{\rm YM} = -\frac{1}{4g^2} \int d^4x \,  F_{a\mu\nu} F_a^{\mu\nu} \, ,
\eea
where $F_{a\mu\nu}$ denotes the YM field strength 
$F_{a\mu\nu}=\partial_\mu A_{a\nu} - \partial_\nu A_{a\mu}
+ f_{abc} A_{b\mu} A_{c\nu}$. On the other hand, 
$S_{W}[A,\rho]$ describes
the gauge-invariant interaction between the colour sources
$\rho$ and the gauge field $A$ built from the 
contour temporal Wilson line $W_C(\vec{x})$, namely
\bea
S_W[A,\rho] = \frac{i}{g} \int d^3\vec{x} ~ {\rm Tr}
[ \rho(\vec{x}) W_C(\vec{x}) ] \, .
\eea
%
We use light-cone coordinates $x^\mu = (x^+,x^-,{\bf x})$ with
$x^\pm = (x^0 \pm x^3)/\sqrt{2}$, ${\bf x} = (x^1,x^2)$ and
$\vec{x} = (x^-, {\bf x})$. 
The contour $C$ is obtained in the limit $x_0^+ \rightarrow -\infty$
and $x^+_f \rightarrow +\infty$ of the
Schwinger-Keldysh contour in the complex time plane
$C_+ \cup C_-$, where $C_+$ is the path along the real time axis
from $x_0^+$ to $x_f^+$ and
$C_-$ is the set of points with a small
imaginary part $z=x^+ - i\eta$, with $\eta \rightarrow 0_+$
and $x^+$ running backwards from $x_f^+$ to $x_0^+$.

At leading order $S_W$ yields a coupling $\rho^a(\vec{x}) A_a^-(\vec{x})$, where $\rho^a$ is the plus component of the fast color current (and the only non-vanishing). Notice that $\rho^a$ is $x^+$-independent, i.e. it represents 
fast static color charges.

The only component of the gauge field that couples to the color sources is $A^-$. Its equation of motion $\frac{\delta S}{\delta A^-} = 0$ yields
in matrix notation
\bea
D_\nu[A]F^{\nu\mu} = \delta^{\mu +} W_{x^+,-\infty}(x) \rho(\vec{x}) W^\dagger_{x^+,-\infty}(x)
\label{eom}
\eea
where we have set $A_\mu = A_{a\mu} T_a$, $\rho=\rho_a T_a$ and
$D_\mu \Phi = \partial_\mu \Phi - i [A,\Phi]$ for a generic field $\Phi=\Phi_a T_a$ in the SU(N) representation spanned by the generators $T_a$.
By keeping the leading $A^-$-independent term in the R.H.S. of eq.(\ref{eom}), 
one can choose a solution where $A^i$ is pure gauge. In the Coulomb gauge
$\partial_l A^l =0$ 
the only non-vanishing component is $A^+$ and 
eq.(\ref{eom}) then reduces to the Poisson equation
\bea
- \nabla^2_{\bf{x}} A^+(\vec{x}) = \rho(\vec{x}) \, ,
\label{poisson}
\eea 
which fixes the gluon field configuration in the presence of the fast
color charges $\rho$. 

\section{BRST Symmetry and Gluon Layers}

In the CGC approach one would like to compute the radiative corrections induced by the exchange of semifast quantum gluons with the classical solution $A^+$
fulfilling eq.(\ref{poisson}).
For that purpose the gluon field $A$ is split into three components as follows:
\bea
A_\mu = \delta A_\mu + a_\mu + \hat A_\mu 
\label{split}
\eea
The background field $\hat A_\mu$ describes the fast configuration
fulfilling eq.(\ref{poisson}) and has support only in the fast
region $|p^+|\geq\Lambda$. The semi-fast gluons $a_\mu$ have support
in the region $b \Lambda < |p^+| < \Lambda$ and have to be integrated out
in the one-step quantum evolution. Finally, $\delta A_\mu$ are the soft modes
at $|p^+|\leq b \Lambda$, so that the full background configuration around which the semifast expansion takes place is $\delta A + \hat A$. Notice that this
is not a stationary point of the YM action.

In order to carry out the integration over $a_\mu$, one needs to fix the gauge. The usual choice is to adopt the light-cone gauge condition $n^\mu A_\mu =0$, while keeping the Coulomb gauge for $\delta A_\mu$.

Once the gauge-fixing is performed, gauge invariance of the classical theory is lost and physical unitarity cancellations are provided by the ST identity, associated with the BRST symmetry of the model. 
In the CGC approach one has in addition the presence of a background source $\hat A_\mu$. The latter is controlled by using an extended BRST symmetry and implementing a canonical transformation that fixes uniquely the dependence on the background $\hat A_\mu$, along the lines of~\cite{Binosi:2012st}.
Consequently one sets $s \hat A_{a\mu} = \Omega_{a\mu}, s\Omega_{a\mu}=0$, where
$\Omega_{a\mu}$ is the BRST partner of the background. The derivative of the effective action w.r.t $\Omega_{a\mu}$ can be understood as the generating functional of the canonical transformation which controls the dependence on $\hat A_\mu$.
Moreover, the BRST transformation of the full gauge field $A_\mu$ is bound to be
$sA_{a\mu} = \partial_\mu c_a + f_{abc} A_{b\mu} c_c$, where $c_a$ is the ghost.
While the BRST transformation of $A_{a\mu}$ and of $\hat A_{a\mu}$ are fixed,
there is an ambiguity in the choice of how $\delta A_{a\mu}$ and $a_{a\mu}$ must transform, since only the sum of their BRST transformation is constrained.
This ambiguity is resolved by noticing that one wishes
to preserve gauge invariance of $S_W[\delta A,\rho]$~\cite{Hatta:2005rn}, so $\delta A$ must transform as a gauge connection, i.e. $s \delta A_{a\mu} = \partial_\mu c_a + f_{abc} \delta A_b c_c$ and thus finally $s a_{a\mu} = f_{abc} (a_{b\mu} + \hat A_{b\mu}) c_c - \Omega_{a\mu}$.
 
Also the colour sources $\rho$ have to be split according to 
$\rho = \delta \rho + \hat \rho$, where $\hat \rho$ are the classical
fast colour sources and $\delta \rho$ are the quantum corrections
to the classical approximation. 
The full source $\rho_a$ transforms in the adjoint representation, 
i.e. $s \rho_a = f_{abc} \rho_b c_c$. On the other hand,
by denoting with $\alpha$ the plus component of the background
$\alpha \equiv \hat A^+$, from eq.(\ref{poisson}) one gets
\be
\hat \rho(\vec{x}) = - U(x) \nabla^2_{\bf{x}} \alpha(\vec{x}) U^\dagger(\vec{x}) \, , \qquad U^\dagger(\vec{x}) = {\rm P} \Big [
i \int_{-\infty}^{x^-} \mathrm{d}z^- \alpha(z^-,\vec{x}) \Big ]
\label{eomU}
\ee
from which one obtains the BRST transformation of $\hat \rho$, induced
by the BRST transformation of $\hat A$:
\be s\hat\rho = - \int \mathrm{d}^4z
\, \Omega_a^+(z) \frac{\delta}{\delta \alpha^a(z)} \hat \rho(\alpha) \, .
\label{srho}
\ee

\section{Slavnov-Taylor identity}

The BRST transformation is nonlinear and therefore a suitable set
of external sources, known as antifields~\cite{Gomis:1994he}, is required in order to formulate its quantum version in the
form of the ST identity. For each quantum field $\varphi$
with a nonlinear BRST variation
one introduces a source $\varphi^*$ with opposite statistics 
and ghost charge ${\rm gh}(\varphi^*) = -1 - {\rm gh}(\varphi)$.
The BRST variation of $\varphi$
is coupled  to $\varphi^*$ 
in the antifield-dependent part $S_{\rm AF}$ of the action, namely
%
$S_{\rm AF} = \int \mathrm{d}^4x ~ \varphi^*(x) ~ s\varphi(x) \, .$
%
Then the full vertex functional $\G$ of the CGC obeys the ST identity
\bea
& {\cal S}(\G) = \int \mathrm{d}^4x \, 
\Big [  \Omega_{a\mu}(x) \frac{\delta \G}{\delta \hat A_{a\mu}(x)}
+ \frac{\delta \G}{\delta (\delta A^{*a}_\mu(x))} 
  \frac{\delta \G}{\delta (\delta A^a_\mu(x))} 
+  \frac{\delta \G}{\delta a^{*a}_\mu(x)} 
  \frac{\delta \G}{\delta a^a_\mu(x)}
+ \frac{\delta \G}{\delta c^*_a(x)}
  \frac{\delta \G}{\delta c_a(x)} \nonumber \\
& \qquad \qquad + \frac{\delta \G}{\delta (\delta \rho^{*a}(x))} 
      \frac{\delta \G}{\delta (\delta \rho^a(x))} +
    b^a(x) \frac{\delta \G}{\delta \bar c^a(x)} \Big ] = 0 \, ,
\label{sti}
\eea
where $b^a$ is the usual Nakanishi-Lautrup field implementing the
gauge-fixing constraint and $\bar c_a$ is the antighost.

After the one-step quantum evolution, 
both the background quantum-splitting in
eq.(\ref{split}) and the classical YM equation of motion
(\ref{poisson}) are  in general deformed.
If this were the case, as for instance if one
quantizes YM theory around an instanton background~\cite{Binosi:2012pd},
the assumption that the YM equation of motion still holds
and that the classical
background splitting is still fulfilled would be violated.
This in turn would prevent
the identification of  the momenta of the updated classical probability
distribution $W_{b\Lambda}$ (at the energy scale $b\Lambda$)
with the quantum correlators of the
color charge fluctuations $\delta \rho$, and consequently
the EFT approach of the CGC could not be pursued in a mathematically
consistent fashion.

However it turns out that indeed the ST identity (\ref{sti}) takes care of all the magic: it guarantees that the background-quantum splitting still
holds in its classical form also after the one-step quantum evolution and moreover that the functional form of the equation of motion for the background remains unaltered. Let us now sketch the proof of both results.

\subsection{Background deformation}

As a consequence of the extended ST identity, 
it can be proven~\cite{Binosi:2011ar}-\cite{Binosi:2012st}
 that the classical background configuration is deformed
as
\bea
\hat A_{a\mu} \rightarrow \hat A_{a\mu} + \G_{\Omega_{a\mu} \delta A^*_{b\nu}} \hat A_{b\nu} + \dots
\eea
where the dots denote higher orders in $\hat A$.
The deformation function $\G_{\Omega_{a\mu} \delta A^*_{b\nu}} \equiv
\frac{\delta^2 \G}{\delta \Omega_{a\mu} \delta(\delta A^*)_{b\nu}}$ is in general
non-vanishing. However in the light-cone gauge for the semi-fast modes
there are no interactions
between the source $\Omega$ and the quantum gluons $a$ and thus
$\G_{\Omega_{a\mu} \delta A^*_{b\nu}}$ turns out to be zero (even
at $\hat A \neq 0$). Therefore
the background (which is kept in the Coulomb gauge) is not deformed. 
Incidentally we notice that this justifies the expansion around the
 configuration of the classical  background plus $\delta A$ (at fixed $\delta A$) although this is not a minimum of the action and 
therefore one is not carrying out a saddle point approximation.

\subsection{Background equation of motion}

We now take a derivative w.r.t. $\Omega_{a\mu}$ of eq.(\ref{sti})
and then consider the sector at $c=\Omega=b=0$. 
Since $\G_{\Omega_{a\mu} \delta A^*_{b\nu}}$ is zero, we obtain the simpler
equation
\be
\frac{\delta \G}{\delta \hat A^a_\mu(x)} = 
-\int \mathrm{d}^4z \frac{\delta^2 \G}{\delta \Omega^a_\mu(x) 
\delta (\delta \rho^{*b}(z))} \frac{\delta \G}{\delta(\delta \rho^b(z))} \, .
\ee
This is the equation of motion for the background which holds when
quantum corrections are taken into account. Since $\delta \rho$
does not enter into the gauge-fixing, it follows that
the deformation function $\G_{\Omega^a_\mu \delta \rho^*_b} \equiv
\frac{\delta^2 \G}{\delta \Omega^a_\mu \delta \rho^*_b}$ stays classical,
i.e. $\G_{\Omega^a_\mu \delta \rho^*_b} = \G^{(0)}_{\Omega^a_\mu \delta \rho^*_b}$.
In the leading logarithimic approximation only the first term of the
BRST transformation of $\hat \rho$ in eq.(\ref{srho}) is important and one finds
$\G^{(0)}_{\Omega^\mu_a(x) \delta \rho^*_b(y)} = \nabla^2_{\bf x} \delta^{(4)}(x-y)\delta_{ab}$,
i.e. one recovers the classical equation of motion (\ref{eomU}) when one sets $\alpha=0$
in the Wilson line $U$.

\section{Evolution equation}

By the above results we know that the dependence on the 
scale $b$ cannot enter via the deformation function $\G_{\Omega \delta A^*}$
and $\G_{\Omega \delta \rho^*}$ (since the first is zero and the latter
is purely classical). Hence we can write  the following
RG equation for the theory in the sector $\delta A-\delta \rho$:
\bea
\frac{\partial\Gamma}{\partial b}&=&\int\!\sum_{n=0}^{\infty}\sum_{m=0}^\infty\frac1{{n!m!}}\widehat{A}^{a_1}_{\mu_1}(x_1)\cdots\widehat{A}^{a_n}_{\mu_n}(x_n)\nonumber \\
&\times&
\left(\widehat{\rho}^{\,d_1}+\Gamma^{(0)}_{\Omega^{d_1}_{\mu_1}\, \delta \rho^{*}_{e_1}}\widehat{\rho}^{\,e_1}+\cdots\right)(y_1)\cdots\left(\widehat{\rho}^{\,d_m}+\Gamma^{(0)}_{\Omega^{d_m}_{\mu_m}\, \delta \rho^{*}_{e_m}}\widehat{\rho}^{\,e_m}_{\nu_m}+\cdots\right)(y_m)\nonumber \\
&\times&\left.\frac{\partial}{\partial b}\frac{\delta^{m+n}\Gamma}{\delta(\delta A^{a_1}_{\mu_1}(x_1))\cdots\delta(\delta A^{a_n}_{\mu_n}(x_n))\delta(\delta \rho^{d_1}(y_1))\cdots\delta(\delta \rho^{d_m}(y_m))}\right\vert_{\delta A=\delta\rho=\widehat{A}=\widehat{\rho}=0}
\label{eveq}
\eea   
This equation provides the RG evolution equation for the
EFT of the CGC in a very general setting.
We stress that it is not yet the JIMWLK evolution equation, since one has to identify
the correlators of $\delta \rho$ in the effective field theory with the
momenta of the weight fucntion $W_{b\Lambda}$ at the new energy scale
$b\Lambda$. After this identification, 
one can derive from eq.(\ref{eveq}) the functional
differential operator controlling the transition from $W_\Lambda[\rho]$
to $W_{b\Lambda}[\rho]$.


The ensuing algebraic derivation as well as the details of the analysis
sketched here will be presented elsewhere~\cite{Binosi:2014xua}.

\end{document}